\documentclass[aps,prc,twocolumn,nofootinbib,preprintnumbers,superscriptaddress,floatfix,showpacs]{revtex4-1}
\usepackage{amsmath,graphicx}

\usepackage{epsfig}
\usepackage{dcolumn}
\usepackage[latin1]{inputenc}

\begin{document}

\title{Hydrodynamic predictions for 5.02 TeV Pb-Pb collisions}

\author{Jacquelyn Noronha-Hostler}
\affiliation{Department of Physics, Columbia University, New York, 10027, USA}
\author{Matthew Luzum}
\affiliation{Instituto de F\'{i}sica, Universidade de S\~{a}o Paulo, C.P.
66318, 05315-970 S\~{a}o Paulo, SP, Brazil}
\affiliation{Departamento de F\'isica de Part\'iculas and IGFAE,
Universidade de Santiago de Compostela, E-15706 Santiago de
Compostela, Galicia-Spain}
\author{Jean-Yves Ollitrault}
\affiliation{Institut de physique th\'eorique, Universit\'e Paris Saclay, CNRS, CEA, F-91191 Gif-sur-Yvette, France}

\begin{abstract}
We make predictions for momentum-integrated elliptic and triangular flow as well as mean transverse momentum for 5.02 TeV Pb-Pb collisions, as planned at the Large Hadron Collider.  We use hydrodynamic calculations to predict the change of these observables as the center-of-mass collision energy evolves from 2.76 TeV to 5.02 TeV per nucleon pair.  By using previously measured values as a baseline, we are able to make a robust prediction without relying on a particular model for initial conditions and without precise knowledge of medium properties such as viscosity.  Thus, though the predicted changes are small, they can provide a significant test of the current hydrodynamic picture of heavy-ion collisions.  
\end{abstract}

\maketitle

\section{Introduction}
The heavy ion program of Run 2 of the Large Hadron Collider (LHC) is scheduled to commence in late 2015 with collisions of lead ions at an energy of 5.02 TeV per nucleon pair.  This follows lower energy collisions of 2.76 TeV in Run 1, as well as collisions of various ions at the Relativistic Heavy-Ion collider up to 200 GeV per nucleon pair.
During this time, a consensus picture has emerged of the collision
system evolving according to the equations of relativistic viscous
fluid dynamics~\cite{Heinz:2013th}.  Many hydrodynamic calculations
have been performed, showing remarkable agreement with a wide variety
of
observables~\cite{Broniowski:2008vp,Song:2011hk,Gardim:2012yp,Gale:2012rq,Qiu:2012uy},
including a number of predictions~\cite{Luzum:2009sb,Alver:2010dn,Bhalerao:2011ry,Bozek:2011if,Nagle:2013lja,Yan:2013laa,Kozlov:2014fqa} made before the measurements were
performed~\cite{Aamodt:2010pa,ALICE:2011ab,Chatrchyan:2013kba,CMS:2012qk,Khachatryan:2015waa,Adare:2015ctn,Aad:2014lta}.
The new heavy-ion run presents an opportunity to further test this picture in a new energy regime. 

Despite this success of hydrodynamic models, there remains significant uncertainties.  In particular, the initial, non-equilibrium stages of the collision are not well understood, and similarly for temperature-dependent transport coefficients. Generically,  a minimum in $\eta/s(T)$ is expected near the cross-over temperature \cite{Hirano:2005wx,Csernai:2006zz} both from the hadron gas phase \cite{NoronhaHostler:2008ju,NoronhaHostler:2012ug} and the QGP phase  \cite{Xu:2014tda,Wesp:2011yy,Hidaka:2009ma,Christiansen:2014ypa,Ozvenchuk:2012kh,Cremonini:2012ny}, but the exact location, magnitude, and slope of that minimum is unknown.   Multiple models for initial conditions~\cite{Luzum:2008cw,Retinskaya:2013gca} and multiple parameterizations of shear viscosity~\cite{Gale:2012rq,Niemi:2015qia}, for example,  are able to give a good description of data.   

In addition to uncertainties in the theory, there are systematic uncertainties in the measurements.  Even if a calculation fits all the data within the experimental error bars, which is rarely the case, one clearly cannot trust any particular calculation to be correct with a precision better than these error bars. Due to these various uncertainties, there is a limit to the precision with which one can trust a prediction for a new collision system, made using a particular model of initial conditions and a single choice for medium properties and freeze-out prescription.

However, we argue that one can actually make precise and reliable predictions without assuming a particular model for initial conditions or values for medium properties.  Instead of choosing a particular model (with parameters chosen to give a reasonable fit to existing data) and doing a single calculation of the new collision system, the idea is to directly use previously measured values as a baseline and focus on the \textit{change} in observables of a Pb-Pb collision system as the energy is increased from 2.76 TeV to 5.02 TeV per nucleon pair.

By doing this, the uncertainties are significantly reduced --- not only theoretical uncertainties, but also experimental systematic uncertainties, which will partially cancel for pairs of measurements done at the two collision energies using the same detector and with the same analysis.  As a result, we can do multiple calculations with many initial conditions and model parameters, to obtain a prediction that is more precise and robust.

In the same vein, to make a prediction that is as reliable as possible, we focus on bulk, momentum-integrated observables for unidentified charged hadrons, which have significantly smaller uncertainties than more differential observables or more rare particles.  Specifically, we focus on mean transverse momentum $\langle p_t\rangle$ along with integrated elliptic flow $v_2\{2\}$ and triangular flow $v_3\{2\}$.
\section{Calculations and Results}
We aim to have a prediction that is as model-independent as possible.
One input that is needed is the charged hadron multiplicity at the
increased collision energy.  In the following, we use the parameterization of the 
scaling of multiplicity~\cite{Aamodt:2010pb,ATLAS:2011ag} 
with collision energy $dN/d\eta\sim s^{0.155}$, where the exponent uses the recent measurement 
at 5.02~TeV~\cite{Adam:2015ptt}.

In hydrodynamic calculations, the momentum integrated $v_2$ and $v_3$ can be accurately predicted in any given collision event by measures of the initial spatial anisotropy known as eccentricity $\varepsilon_2$ and triangularity $\varepsilon_3$, respectively.  That is,
\begin{equation}
v_n = \kappa_n \varepsilon_n
\end{equation}
with
\begin{equation}
\varepsilon_n = \frac {|\int d^2r\ r^n e^{in\phi} \rho(\vec r)|} {\int d^2r\ r^n \rho(\vec r)} .
\end{equation}
Here $\rho(\vec r)$ is usually taken to be either energy density or entropy density, with each having roughly equivalent predictive power \cite{Gardim:2011xv}.   In this work we calculate both and verify that the result has a negligible difference.

$\kappa_n=v_n/\varepsilon_n$ represents the hydrodynamic response.  It depends on aspects such as medium properties and the freeze-out prescription, but is considered to be independent of the initial conditions and is therefore the same in every event (within a given centrality class).  Therefore, the change in elliptic and triangular flow can be separately studied as a change in the initial $\varepsilon_n$, compounded by a change in the hydrodynamic response $v_n/\varepsilon_n$.

We address the former by gathering a set of models for initial conditions and calculating the change in $\varepsilon_2$ and $\varepsilon_3$.  Since $v_n\{2\}\sim\sqrt{\langle v_n^2 \rangle}$, we are interested in the change in the root mean square.  Here we investigate the following models --- MC-Glauber~\cite{Miller:2007ri,Alver:2008aq,Rybczynski:2013yba}, MC-KLN~\cite{Drescher:2006ca}, MCrcBK~\cite{ALbacete:2010ad}, and Trento~\cite{Moreland:2014oya}.

\begin{figure}
\includegraphics[width=\linewidth]{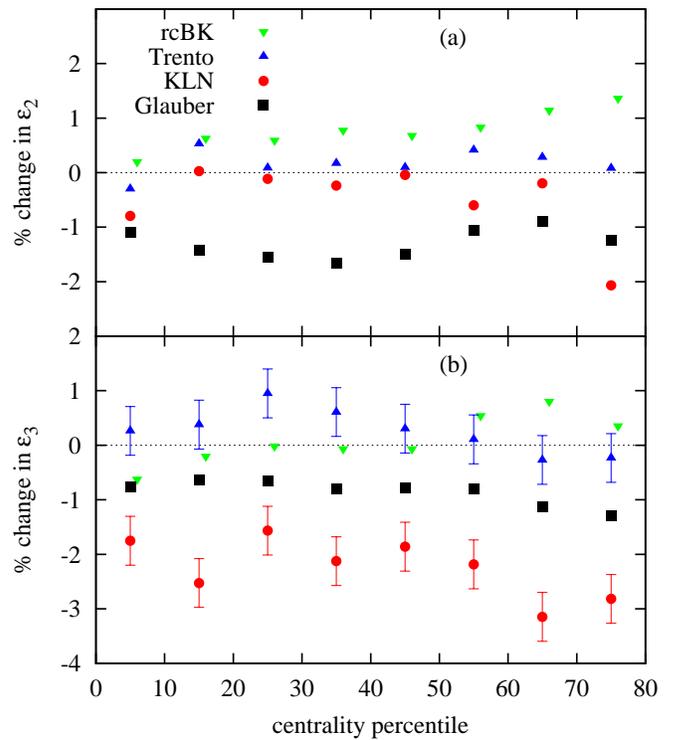}
\caption{\label{ecc}Percent change in rms eccentricity $\varepsilon_2$ (a) and triangularity $\varepsilon_3$ (b) when the collision energy is increased from 2.76 TeV to 5.02 TeV, for several models of initial conditions for heavy-ion collisions --- MC-Glauber~\cite{Alver:2008aq}, Trento~\cite{Moreland:2014oya}, MC-KLN~\cite{Drescher:2006ca}, and MCrcBK~\cite{ALbacete:2010ad}.}
\end{figure}

Each of these models uses the measured nucleon-nucleon inelastic cross
section $\sigma_{\rm inel}$ as input.
This quantity is not measured directly at $2.76$ or $5.02$~TeV. 
At $2.76$~TeV, we use the standard 
value~\cite{ATLAS:2011ag,Abelev:2013qoq} $\sigma_{\rm inel}=64$~mb.
For the extrapolation to $5.02$~TeV, we use 
the parameterization of the total $pp$ cross section by the 
Particle Data Group Collaboration~\cite{Agashe:2014kda} and the 
parameterization of the ratio of elastic to total cross section from
Ref.~\cite{Fagundes:2011hv} \footnotetext{
This parameterization gives 
$\sigma_{\rm inel}=63.4$, $69.2$ and $72.5$~mb at $\sqrt{s}=2.76$,
$5.02$ and $7$~TeV, respectively. 
The latter value is in agreement with the measurement $72.9\pm 1.5$~mb 
by the TOTEM collaboration~\cite{Antchev:2013iaa}. We round up $63.4$
and $69.2$ to $64$ and $70$, respectively.}
and choose $\sigma_{\rm inel}=70$~mb. 

For the Glauber model, we use the PHOBOS Monte Carlo Glauber
\cite{Alver:2008aq} v1.0 to calculate eccentricities.  We use the binary collision fraction from~\cite{Abelev:2013qoq}.  The only change
with collision energy is the inelastic cross section, as described
above.   

Trento \cite{Moreland:2014oya} is a phenomenological model with
parameters that allow one to smoothly interpolate between various
types of initial conditions.  For example, one limit corresponds to a
Glauber model with participant scaling.  
Here, we choose the set of parameters that best reproduce multiplicity distributions  in Pb+Pb, p+Pb, and p+p collisions at the LHC, and which approximately correspond to eccentricities from the IP-Glasma model: $p=0$, $k=1.4$ \cite{Moreland:2014oya}.

MC-KLN \cite{Drescher:2006ca} is a model based on saturation physics that calculates gluon production using the $k_t$ factorization formula and an ansatz for the unintegrated gluon distribution. We calculate using default parameters from the latest version mckln-3.52.  

Lastly, we use the MCrcBK model \cite{Dumitru:2012yr}, calculated with default parameters from mckt-v1.32.  It is similar to MC-KLN in that it uses the $k_t$ factorization formula as a starting point.  However, instead of the KLN ansatz for the unintegrated gluon distributions, they are calculated from the running coupling BK equation.

The results from all models are shown in Figure \ref{ecc}.  Over all centralities and every model, the change from 2.76 TeV to 5.02 TeV is between -2\% and 2\% for rms $\varepsilon_2$ and between -3\% and 1\% for $\varepsilon_3$.  

To investigate the change in the hydrodynamic response, we perform viscous hydrodynamic calculations using the 2+1 relativistic viscous hydrodynamical code, v-USPhydro \cite{Noronha-Hostler:2013gga,Noronha-Hostler:2014dqa}.  v-USPhydro uses Smoothed Particle Hydrodynamics, a Lagrangian method to solve the equations of motion on an event-by-event basis, where the smoothing parameter is set to $\lambda=0.3$ fm for all calculations \cite{Noronha-Hostler:2015coa}.  

We start by generating events from a Monte Carlo Glauber model, with a scale factor chosen for each set of parameters to match the charged hadron multiplicity at 2.76 TeV.  Then we scale each initial condition by a constant factor to calculate the hydrodynamic response at 5.02 TeV.   In this way, the geometry of each event is fixed, and we can isolate the change in the hydrodynamic response $v_n/\varepsilon_n$ and in $\langle p_t\rangle$.  The actual eccentricity distribution of the Glauber model used, is therefore irrelevant.  We use the lattice-based equation of state EOS S95n-v1 from \cite{Huovinen:2009yb},
and calculate the change of observables using all charged hadrons including all resonance decays~\cite{Kolb:2000sd,Kolb:2002ve}.
The following $p_t$-integrated results correspond to the range $p_t>0.2$ GeV, but predictions for other relevant ranges are presented in \ref{appendix}.

\begin{figure}[t]
\includegraphics[width=\linewidth]{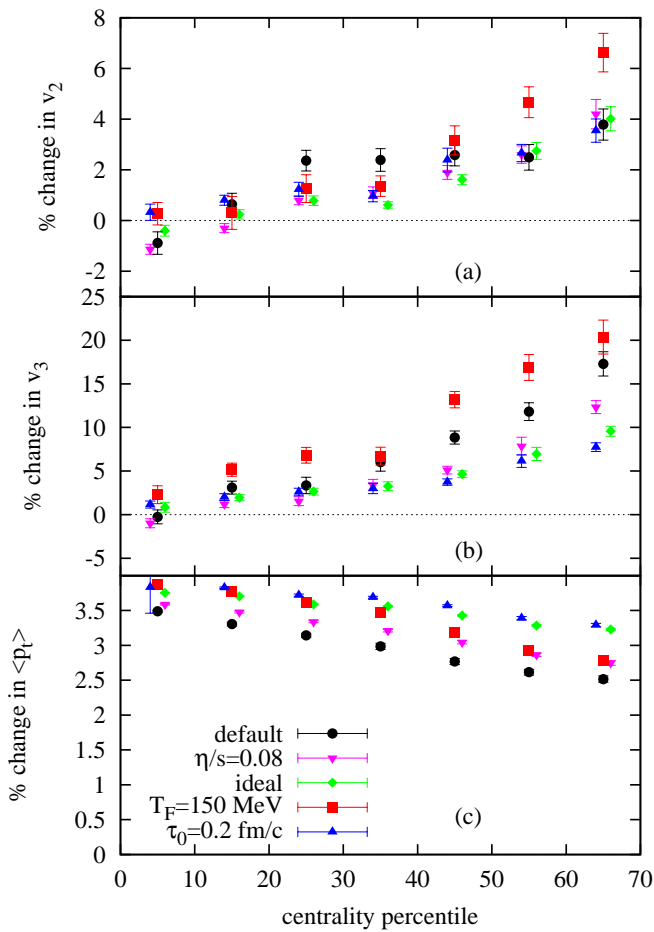}
\caption{\label{response}Percent change in the hydrodynamic response $v_2/\varepsilon_2$ (a) and $v_3/\varepsilon_3$ (b), as well as mean transverse momentum $\langle p_t\rangle$ (c) when the collision energy is increased from 2.76 TeV to 5.02 TeV, for several sets of hydrodynamic parameters.  Error bars represent statistical uncertainty from the finite number of generated events.  See text for details.}
\end{figure}

We choose for a default set of parameters $\tau_0=0.6$~fm$/c$, $T_{FO}=130$ MeV, and a temperature-dependent sheer viscosity $\eta/s(T)$ as labelled `param1' in \cite{Niemi:2015qia} (converted to a chemical equilibrium Equation of State as in Fig. 23 from that reference).   The change in $v_2\{2\}$, $v_3\{2\}$, and $\langle p_t\rangle$ for Run 2 relative to Run 1 are shown in Fig.~\ref{response}.  In general, the predicted changes are at the several percent level, with the mean transverse momentum increasing by approximately 3\% across all centralities.  
Recall that a non-trivial prediction from hydrodynamics is that $\langle p_t\rangle$ depends little on centrality~\cite{Bozek:2012fw}, and so it is unsurprising that the \textit{change} also depends little on centrality.
The mean transverse momentum is driven by 
by the equation of state both in ideal~\cite{Blaizot:1987cc} and
viscous~\cite{Bozek:2012fw} hydrodynamics. More precisely, it scales
like the energy per particle at the time when transverse flow builds
up. The multiplicity near midrapidity $N_{\rm ch}$ is proportional to
the entropy, therefore 
$\langle p_t\rangle$ scales like $\epsilon/s$, where $\epsilon$ and
$s$ denote the energy and entropy density
respectively~\cite{Ollitrault:1991xx}. Hence the relative increase of
$\langle p_t\rangle$ is 
\begin{equation}
\label{deltapt}
\frac{d\langle p_t\rangle}{\langle
  p_t\rangle}=\frac{d\epsilon}{\epsilon}-\frac{ds}{s}=\frac{P}{\epsilon}\frac{dN_{\rm  ch}}{N_{\rm ch}}, 
\end{equation}
where $P$ is the pressure, and we have used the thermodynamic relations
$d\epsilon=Tds$, $\epsilon+P=Ts$, $ds/s=dN_{\rm  ch}/N_{\rm  ch}$.
The temperature when transverse flow builds up at the LHC is in the
range $200-250$~MeV~\cite{Shen:2011eg}. 
In this temperature range, 
$P\simeq\epsilon/6$~\cite{Huovinen:2009yb}, and Eq.~(\ref{deltapt}) 
explains why a 20\% increase in multiplicity corresponds to a 3\% 
increase in $\langle p_t\rangle$. Our prediction, if confirmed by
experimental data, thus provides a direct experimental verification of
the QCD equation of state.

While the predicted increase in $\langle p_t\rangle$ depends little on centrality, elliptic and triangular flow have the largest increase  in peripheral collisions.  This is also expected generically.  Note that anisotropic flow is generated as a response to a spatial anisotropy.  In a given collision event, this spatial anisotropy decreases with time, and the generation of anisotropic flow therefore slows.  Peripheral events, with the shortest lifetime, are still rapidly generating $v_n$ when they freeze out, and therefore generate significantly more when the lifetime is increased.  On the other hand, central collisions already have a long lifetime at 2.76 TeV, and have therefore essentially saturated the anisotropic flow.  In fact, the spatial anisotropy can eventually go through zero and change sign, resulting in a slight \textit{decrease} in $v_n$ with extra lifetime.    From Fig.~\ref{response}, one can see that this is indeed our prediction for $v_2$ in the most central collisions. 

\begin{figure}[t]
\includegraphics[width=\linewidth]{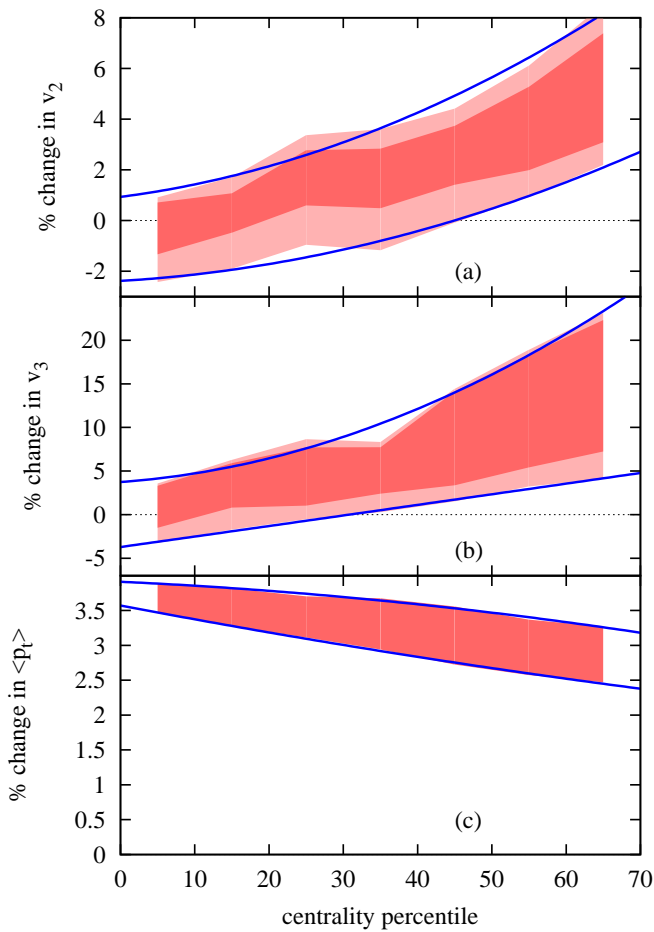}
\caption{\label{prediction}Predicted percent change in $v_2\{2\}$ (a), $v_3\{2\}$ (b), and $\langle p_t\rangle$ (c) when the collision energy is increased from 2.76 TeV to 5.02 TeV.  The dark band indicates the predicted change only from the hydrodynamic response, which is the dominant contribution to the error band, while the light band in the upper two plots includes the effect from a change in initial eccentricity.  The solid lines represent a polynomial fit to the limits of the error band, which can be used to interpolate our prediction to any particular centrality.}
\end{figure}

While this set of hydrodynamic parameters can give a reasonable fit to existing data given a judicious choice of initial conditions, there is uncertainty in their values.  Because of this, we vary each parameter, in order to get an idea of the robustness of our prediction.  In each case, we rescale the initial conditions in order to keep the final multiplicity fixed.  

Fig.~\ref{response} shows the result when we separately change the shear viscosity from our fairly strong default temperature dependence, to constant values of $\eta/s=0.08$~\cite{Kovtun:2004de} and $\eta/s=0$.  
The latter choice is motivated by the observation that there is no lower bound on $\eta/s$ from experimental data 
alone~\cite{Gardim:2012yp}.
In addition, we vary the freeze-out temperature to $T_{FO}=150$ which was found to be the highest value able to fit
$p_t$ spectra at the LHC~\cite{Bozek:2012qs}. 
We vary the  initial time to $\tau_0=0.2$~fm$/c$  to mimic the effect of initial transverse flow
\cite{Vredevoogd:2008id,Liu:2015nwa,vanderSchee:2013pia}. 

Note that, while these parameters can have a very significant affect on the absolute values of $v_2\{2\}$, $v_3\{2\}$, and $\langle p_t\rangle$, the affect of changing the parameters tends to be similar at each collision energy, and therefore there is a much smaller affect on the \textit{difference}.   In fact, it's possible that not all of these parameter sets would be able to fit data even at 2.76 TeV,  but these results should give a conservative idea of the size of uncertainty in our prediction.

We can see that the uncertainty in our prediction for $\langle p_t\rangle$ is largely due to the uncertainty in $\eta/s$, while for $v_2$ and $v_3$ it is more sensitive to freeze out.  The uncertainty in the change in initial eccentricity appears to be much smaller than in the hydro response.

In Fig.~\ref{prediction} we combine the changes in initial spatial anisotropy and the hydrodynamic response, using the range of results as an estimate of systematic uncertainty, to show our overall predictions for the change in each measured quantity.

We note that we have not investigated the effect of bulk viscosity, which could potentially increase our error band.   However, since bulk viscosity is expected to decrease quickly at high temperature, we expect that its effect should decrease with collision energy.

In a recent preprint \cite{Niemi:2015voa}, predictions were given using the EKRT model of initial conditions and several choices of $\eta/s(T)$.  For the parameterizations  of $\eta/s(T)$ that work best across various energies, they fall within our band of theoretical uncertainties.  However, the two parameterizations that fail to fit RHIC data and correlations of event-plane angles \cite{Niemi:2015qia} would predict a larger anisotropic flow in central collisions than our result.

Finally, we take the results from existing measurements, and scale up by our predicted change, to  give an absolute prediction for the values of $v_2\{2\}$ and $v_3\{2\}$ for various analyses from ALICE, CMS, and ATLAS.  The results are shown in \ref{appendix}.  

\section{Conclusions}
We have presented predictions for the upcoming heavy-ion run at the LHC.  By using previous measurements at lower energy as a baseline, we are able to make precise predictions for the evolution of observables as the collision energy is increased from 2.76 TeV to 5.02 TeV, and which can be tested with significant precision by performing the same experimental analyses at both collision energies, resulting in a reduced systematic uncertainty on the change with energy.

Further, we make our predictions as model-independent as possible by focussing on specific observables and doing numerous calculations for various models and parameters to make a robust prediction.  Specifically, we choose $\langle p_t\rangle$ because it is insensitive to model details, and momentum-integrated $v_2\{2\}$ and $v_3\{2\}$, because they have much smaller uncertainty than differential measurements, and because these particular harmonics allow for a linear response analysis.

Compared to the lower energy LHC measurement, we predict that $\langle p_t \rangle$ will increase between 2.5\%-3.5\% with the largest increase in central collisions, but little centrality dependence overall.  $v_2\{2\}$ and $v_3\{2\}$ will see the largest increases in peripheral collisions, of at least several percent, while in central collisions $v_3\{2\}$ and $v_2\{2\}$ will see little change.

These predictions provide an opportunity to precisely test the hydrodynamic paradigm of heavy-ion collisions.  Any deviation of measurements from these predictions would highlight possible gaps in our current understanding, while the exact measured value within the predicted range could determine features such as the temperature dependence of viscosity, bulk viscous effects, and hadronic physics at freeze out. 

\begin{acknowledgments}
We thank the Institute for Nuclear Theory at the University of Washington for its hospitality and the Department of Energy for partial support during the initiation of this work.
ML was supported by Marie Curie Intra-European Fellowship for Career Development grant FP7-PEOPLE-2013-IEF-626212.  JNH acknowledges
support from the US-DOE Nuclear Science Grant No.~DE-FG02-93ER40764.
 JYO thanks the departamento de f\'\i sica of Universidad de Buenos Aires for hospitality.
 \end{acknowledgments}

\appendix
\section{Prediction Tables}
\label{appendix}
We present tables of predictions corresponding to the transverse momentum range of measurements at 2.76 TeV from ALICE (Table \ref{ALICE}), CMS (Table \ref{CMS}) and ATLAS (Table \ref{ATLAS}), obtained by taking our prediction for the fractional change in each observable from 2.76 to 5.02 TeV and multiplying by the respective measured value from LHC Run 1.  The $p_t$ ranges and corresponding measurements are $p_t\geq 0.2$ GeV for ALICE \cite{ALICE:2011ab}, $p_t\geq 0.3$ for  CMS \cite{Chatrchyan:2012ta} and $p_t\geq 0.5$ for ATLAS \cite{ATLAS:2012at}. 

\begin{table}
\centering
\begin{tabular}{ c | c | c  }
 & \multicolumn{2}{c}{ALICE ($p_t\geq 0.2$ GeV)}\\[5pt]
\hline
\rule{0pt}{1.0\normalbaselineskip}
Cent. &  $v_2\{2\}$ & $v_3\{2\}$ \\[2pt]
\hline
\rule{0pt}{1.0\normalbaselineskip}
0--5\% &  0.0264 -- 0.0273  &0.0198 -- 0.0213\\
5--10\% & 0.0433  -- 0.0448  &0.0230 -- 0.0247\\
10--20\% &  0.0627 --  0.0651 & 0.0259 -- 0.0279\\
20--30\% &  0.0819 -- 0.0852  &0.0291 -- 0.0316\\
30--40\% & 0.0939 -- 0.0982  &0.0307 -- 0.0339\\
40--50\% & 0.0993 -- 0.1043  &0.0306 -- 0.0346\\
50--60\% & 0.0972 -- 0.1027  &\\
60--70\% & 0.0889  -- 0.0944  &\\
70--80\% & 0.0729 -- 0.0780&
\end{tabular}
\caption{\label{ALICE}Prediction for integrated $v_2\{2\}$ and $v_3\{2\}$, integrated over $p_t\geq 0.2$ GeV, using the measurements in \cite{ALICE:2011ab} as a baseline.
}
\end{table}

\begin{table}
\centering
\begin{tabular}{ c | c | c  }
 & \multicolumn{2}{c}{CMS ($p_t\geq 0.3$ GeV)}\\[5pt]
\hline
\rule{0pt}{1.0\normalbaselineskip}
Cent. &  $v_2\{EP\}$ & $v_3\{EP\}$ \\[2pt]
\hline
\rule{0pt}{1.0\normalbaselineskip}
0--5\% &   0.0264 -- 0.0272 &  0.0203 -- 0.0217\\
5--10\% & 0.0444 -- 0.0460 &  0.0242 --  0.0259\\
10--15\% &  0.0595 -- 0.0618 & 0.0268 --  0.0287\\
15--20\% & 0.0719 -- 0.0748 &  0.0289 -- 0.0311\\
20--25\% & 0.0821 --  0.0856 &  0.0308 --  0.0333\\
25--30\% & 0.0905 -- 0.0946 &  0.0321 -- 0.0349\\
30--35\% & 0.0971 --  0.1018 & 0.0331 --  0.0364\\
35--40\% & 0.1019 --  0.1070 & 0.0341 -- 0.0378\\
40--50\% & 0.1062 --  0.1118 &  0.0342 -- 0.0386\\
50--60\% & 0.1070 --  0.1130 &  0.0323 --  0.0376\\
60--70\% & 0.1006 --  0.1065 &  0.0265 -- 0.0321
\end{tabular}
\caption{\label{CMS}
Prediction for integrated $v_2\{EP\}$ and $v_3\{EP\}$, integrated over $p_t\geq 0.3$ GeV, using the measurements in \cite{Chatrchyan:2012ta} as a baseline.   Since the value of $v_n\{EP\}$ depends on the event plane resolution, these predictions assume the analysis is done in a way to ensure the same resolution at both collision energies.}
\end{table}
\begin{table}
\centering
\begin{tabular}{ c | c | c  }
 & \multicolumn{2}{c}{ATLAS ($p_t\geq 0.5$ GeV)}\\[5pt]
\hline
\rule{0pt}{1.0\normalbaselineskip}
Cent. &  $v_2\{2\}$ & $v_3\{2\}$ \\[2pt]
\hline
\rule{0pt}{1.0\normalbaselineskip}
0--2\% & 0.0274 -- 0.0280 &  0.0247 -- 0.0261\\
2--5\% & 0.0380 -- 0.0390 & 0.0276 -- 0.0292\\
5--10\% & 0.0549 --  0.0568 & 0.0308 -- 0.0326\\
10--15\% & 0.0732 --  0.0764 & 0.0340 -- 0.0362\\
15--20\% & 0.0883 --  0.0928 & 0.0367 -- 0.0392\\
20--25\% & 0.1007 -- 0.1065 &  0.0389 -- 0.0419\\
25--30\% & 0.1112 -- 0.1180 & 0.0410 -- 0.0445\\
30--35\% & 0.1194 -- 0.1271 & 0.0427 -- 0.0469\\
35--40\% & 0.1258 -- 0.1343 & 0.0440 -- 0.0489\\
40--45\% & 0.1302 -- 0.1390 & 0.0450 -- 0.0506\\
45--50\% & 0.1325 -- 0.1415 & 0.0453 -- 0.0518\\
50--55\% & 0.1329 -- 0.1417 & 0.0453 --  0.0525\\
55--60\% & 0.1315  -- 0.1397 & 0.0446 -- 0.0527\\
60--80\% & 0.1251 -- 0.1314 & 0.0451 --  0.0560
\end{tabular}
\caption{\label{ATLAS}Prediction for integrated $v_2\{2\}$ and $v_3\{2\}$, integrated over $p_t\geq 0.5$ GeV, using the measurements in \cite{ATLAS:2012at} as a baseline.
}  %doi:
\end{table}

%doi:

\end{document}